\documentclass[iop,apjl]{emulateapj}
\usepackage{epsfig}
\usepackage{amsmath, amsthm, amssymb}
\usepackage[plainpages=false, colorlinks=true, anchorcolor=blue, linkcolor=blue, citecolor=blue, bookmarks=false, urlcolor=blue]
{hyperref}
\usepackage{color}
\usepackage{microtype}
\usepackage{float}
\usepackage{verbatim}
\usepackage{wrapfig}
\usepackage{graphicx}
\usepackage[caption = false]{subfig}
\usepackage{multirow}
\usepackage{hhline}
\usepackage{booktabs}
\usepackage{array}

\newcommand{\Msun}{$M_\sun$}

\newcommand\Tstrut{\rule{0pt}{3.0ex}}

\newcommand{\shortauth}{S. Swihart}
\newcommand{\slugcom}{Draft version - \today}
\slugcomment{\slugcom}

\lefthead{\sc \footnotesize \slugcom \hfill \shortauth}
\righthead{\sc \footnotesize \slugcom \hfill \shortauth}

\begin{document}

\title{A Multi-wavelength view of the neutron star binary 1FGL J1417.7--4402: A progenitor to canonical millisecond pulsars}

\author{Samuel J. Swihart\altaffilmark{1},
Jay Strader\altaffilmark{1},
Laura Shishkovsky\altaffilmark{1},
Laura Chomiuk\altaffilmark{1},
Arash Bahramian\altaffilmark{1,3},
Craig O. Heinke\altaffilmark{2},
James C.~A.~Miller-Jones\altaffilmark{3},
Philip G. Edwards\altaffilmark{4},
C.~C.~Cheung\altaffilmark{5}
}

\affil{ 
  \altaffilmark{1}{Department of Physics and Astronomy, Michigan State University, East Lansing, MI 48824, USA}\\
  \altaffilmark{2}{Department of Physics, University of Alberta, CCIS 4-181, Edmonton, AB T6G 2E1, Canada}\\
  \altaffilmark{3}{International Centre for Radio Astronomy Research – Curtin University, GPO Box U1987, Perth, WA 6845, Australia}\\
  \altaffilmark{4}{CSIRO Astronomy and Space Science, PO Box 76, Epping, NSW 1710, Australia}\\
  \altaffilmark{5}{Space Science Division, Naval Research Laboratory, Washington, DC 20375, USA}\\
}

\begin{abstract}
The \emph{Fermi} $\gamma$-ray source 1FGL J1417.7--4407 (J1417) is a compact X-ray binary with a neutron star primary and a red giant companion in a $\sim$5.4 day orbit. This initial conclusion, based on optical and X-ray data, was confirmed when a 2.66 ms radio pulsar was found at the same location (and with the same orbital properties) as the optical/X-ray source. However, these initial studies found conflicting evidence about the accretion state and other properties of the binary. We present new optical, radio, and X-ray observations of J1417 that allow us to better understand this unusual system. We show that one of the main pieces of evidence previously put forward for an accretion disk---the complex morphology of the persistent H$\alpha$ emission line---can be better explained by the presence of a strong, magnetically driven stellar wind from the secondary and its interaction with the pulsar wind. The radio spectral index derived from VLA/ATCA observations is broadly consistent with that expected from a millisecond pulsar, further disfavoring an accretion disk scenario. X-ray observations show evidence for a double-peaked orbital light curve, similar to that observed in some redback millisecond pulsar binaries and likely due to an intrabinary shock. Refined optical light curve fitting gives a distance of 3.1$\pm$0.6 kpc, confirmed by a \emph{Gaia} DR2 parallax measurement. At this distance the X-ray luminosity of J1417 is (1.0$^{+0.4}_{-0.3}$) $\times 10^{33}$ erg s$^{-1}$, which is more luminous than all known redback systems in the rotational-powered pulsar state, perhaps due to the wind from the giant companion. The unusual phenomenology of this system and its differing evolutionary path from redback millisecond pulsar binaries points to a new eclipsing pulsar ``spider" subclass that is a possible progenitor of normal field millisecond pulsar binaries.
\end{abstract}

\section{Introduction}
\label{sec:intro}
Millisecond pulsars (MSPs) form when matter and angular momentum are accreted onto a neutron star from a non-degenerate companion, recycling them to very rapid spin periods. These systems are generally observable as low-mass X-ray binaries (LMXBs). The typical MSP recycling process involves accretion from a giant donor star overfilling its Roche-lobe, and ends as the orbital period grows and accretion onto the neutron star eventually stops. The resulting system has a rotationally-powered pulsar primary and a low-mass (0.2--0.3 \Msun) white dwarf companion, with orbital periods ranging from days to weeks \citep{Tauris06}. These types of systems constitute the bulk of the known MSPs in the Galactic field.

The most important advance in expanding the known population of MSP binaries has been multi-wavelength (X-ray, optical, and radio) follow-up observations of \emph{Fermi}-LAT GeV $\gamma$-ray sources, which have led to the discovery of many MSPs in binaries with non-white dwarf companions. With improved statistics, these systems 
have been categorized based on their companion mass: black widows have very light ($M_{c}\lesssim0.10M_{\odot}$), highly ablated companions, while redbacks have non-degenerate, main sequence-like companions ($M_{c}\gtrsim0.2M_{\odot}$) that typically fill a substantial fraction of their Roche lobe \citep{Roberts11}. These systems frequently show lengthy, irregular radio eclipses, making them challenging to discover in normal radio pulsar surveys.

These new binaries show interesting phenomenology, but have also drawn extensive interest as the long sought-after links between MSPs and their LMXB progenitors. In particular, three redbacks have been observed to undergo transitions between an accretion-powered ``disk state'' and a rotationally-powered ``pulsar state'' on timescales of days to months \citep{Archibald09, Papitto13, Bassa14, Roy15, Bogdanov15, Johnson15}. These systems, known as transitional millisecond pulsars (tMSPs), give one view of the end of the MSP recycling process, but also provide essential insights into the physics of low-level accretion onto magnetized compact objects and the interactions between MSPs and their stellar and gaseous environment. 

One challenge in extending these insights to understanding the formation of all MSPs is that all the known tMSPs have short orbital periods ($\lesssim 0.5$ days), and hence will likely not end their lives as the \emph{typical} field MSP binaries, which have degenerate companions and longer orbital periods. The progenitors to such systems are thought to be red giant--neutron star binaries with $> 1$ day periods \citep[e.g.,][]{Tauris99}.

One of the few known systems fitting the latter description is a recent discovery. 1FGL J1417.7--4407 / 3FGL J1417.5--4402 (hereafter J1417) is a Galactic compact binary whose stellar counterpart was first discovered by \citet{Strader15} in an optical survey of unassociated \emph{Fermi}-LAT $\gamma$-ray sources. A variable X-ray source near the center of the \emph{Fermi} error ellipse matches a bright optical source. Optical photometry and spectroscopy revealed that the optical source is a red giant, and modeling its ellipsoidal variations, radial velocities, and rotational velocity showed that the system is an evolved $\sim$0.35 \Msun~late-G/early-K giant in a 5.37 day orbit with a $\sim$2 \Msun, suspected neutron star primary.

This compact object interpretation was confirmed when \citet{Camilo16} found a 2.66 ms radio pulsar at the same location and with the same orbital period and phase as the optical binary discovered by \citet{Strader15}. The long orbital period, giant secondary, and short spin period suggest this system is likely in the late stages of the standard MSP recycling process that will end with a white dwarf companion in a $\gtrsim 8$ day orbit \citep{Podsiadlowski02}, making it the first \emph{typical} MSP binary progenitor identified in the Galaxy.

The associated $\gamma$-ray emission, non-degenerate secondary, and MSP primary make J1417 similar to other redbacks, while its wider orbit, giant companion, and inferred evolutionary track are unique, leading us to suggest it as the first member of the ``huntsman" subclass of MSP binaries. Furthermore, despite the fact that J1417 hosts an MSP, this system shows phenomenology dissimilar to redbacks/tMSPs in their pulsar states: in particular, the optical spectra show double-peaked H$\alpha$ emission lines at nearly all epochs \citep{Strader15}, which is a classic signature of an accretion disk. This emission was observed even during times when the system was observed by \citet{Camilo16} as a radio pulsar, precluding the possibility that a pulsar to disk state change had occurred. 

Another peculiarity is that, at the distance inferred by \citet[][4.4 kpc]{Strader15}, the implied X-ray luminosity was $\sim 1.4\times10^{33}$ erg s$^{-1}$, typical for a redback/tMSP in the disk state but higher than usually observed for a system in the pulsar state. \citet{Camilo16} suggest a nearer distance, which would produce a more typical X-ray luminosity of $\sim 10^{32}$ erg s$^{-1}$, but which still leaves the spectroscopic evidence for an accretion disk unexplained.

To better constrain the properties of J1417 and its connection to redbacks/tMSPs, here we report on new multi-wavelength observations of the system. We describe our newly acquired X-ray, radio, and optical/near-IR observations of the source in \S\ref{sec:xrayobservations}, \S\ref{sec:radioobs}, and \S\ref{sec:optobs}, respectively. In \S\ref{sec:optLCs} we perform detailed modeling of the ellipsoidal light curve, while we attempt to explain the source of the complex H$\alpha$ structures in \S\ref{sec:Halpha}. Concluding remarks are presented in \S\ref{sec:discussion}.

\section{X-ray Observations}
\label{sec:xrayobservations}

\subsection{Chandra ACIS}
\label{sec:chandradata}
\citet{Strader15} analyzed a 2 ks \emph{Chandra}/ACIS-I exposure of the 1FGL field taken in 2011 Feb (ObsID 12843; PI Ricci). The neutron star binary discussed in this work (J1417) was the only significant X-ray source in the field. They found that the X-ray spectrum was well-fit by an absorbed single power-law with photon index $\Gamma$ = 1.32~$\pm$~0.40, but that this could not be distinguished from an absorbed blackbody spectrum (kT = 0.78$_{-0.12}^{+0.16}$ keV) due to the small number of counts. No significant limits could be placed on the source variability. These data sampled orbital phases $\phi = 0.981 - 0.986$.

To achieve improved constraints on the short-term X-ray variability, as well as an improved spectrum, we obtained a single, uninterrupted $\sim$50 ks Chandra exposure of J1417 (ObsID 17786; PI Strader) on 2016 Jun 6 using the ACIS-S3 CCD configured in FAINT mode. To avoid pileup, the CCD was used in a custom timed exposure subarray mode with 256 pixel rows, starting from CCD row 385.

The data were reprocessed, reduced, and analyzed with \texttt{CIAO 4.9} using the \texttt{CalDB 4.7.4} calibration files \citep{Fruscione06}. We found no evidence for background flares \citep[these are thought to come from low-energy solar or geomagnetic protons;][]{Grant02}. The data gave a net exposure of 49771.0 s, covering $\sim$11\% of the orbit from $\phi = 0.86 - 0.97$. We note that here and throughout this paper, we adopt the ephemeris from \citet{Camilo16}, where $\phi = 0$ is the ascending node of the pulsar. We extracted spectra of the source with \texttt{specextract} using a circular region of radius 2.5$\arcsec$, with the background determined from a nearby circular, source-free region with radius 25$\arcsec$, yeilding a net count rate of 60.3 $\pm$ 1.1 cts ks$^{\textrm{-1}}$ (2886 photons). To analyze the spectra using \texttt{Xspec} \citep{Arnaud96}, we binned the data to have at least 30 counts per bin, and assumed \citet{Wilms00} abundances for the photoelectric absorption models throughout.

\begin{figure}[!b]
    \vspace{10pt}
    \centering
    \includegraphics[width=1.0\linewidth,trim={0 0 0 0},clip]{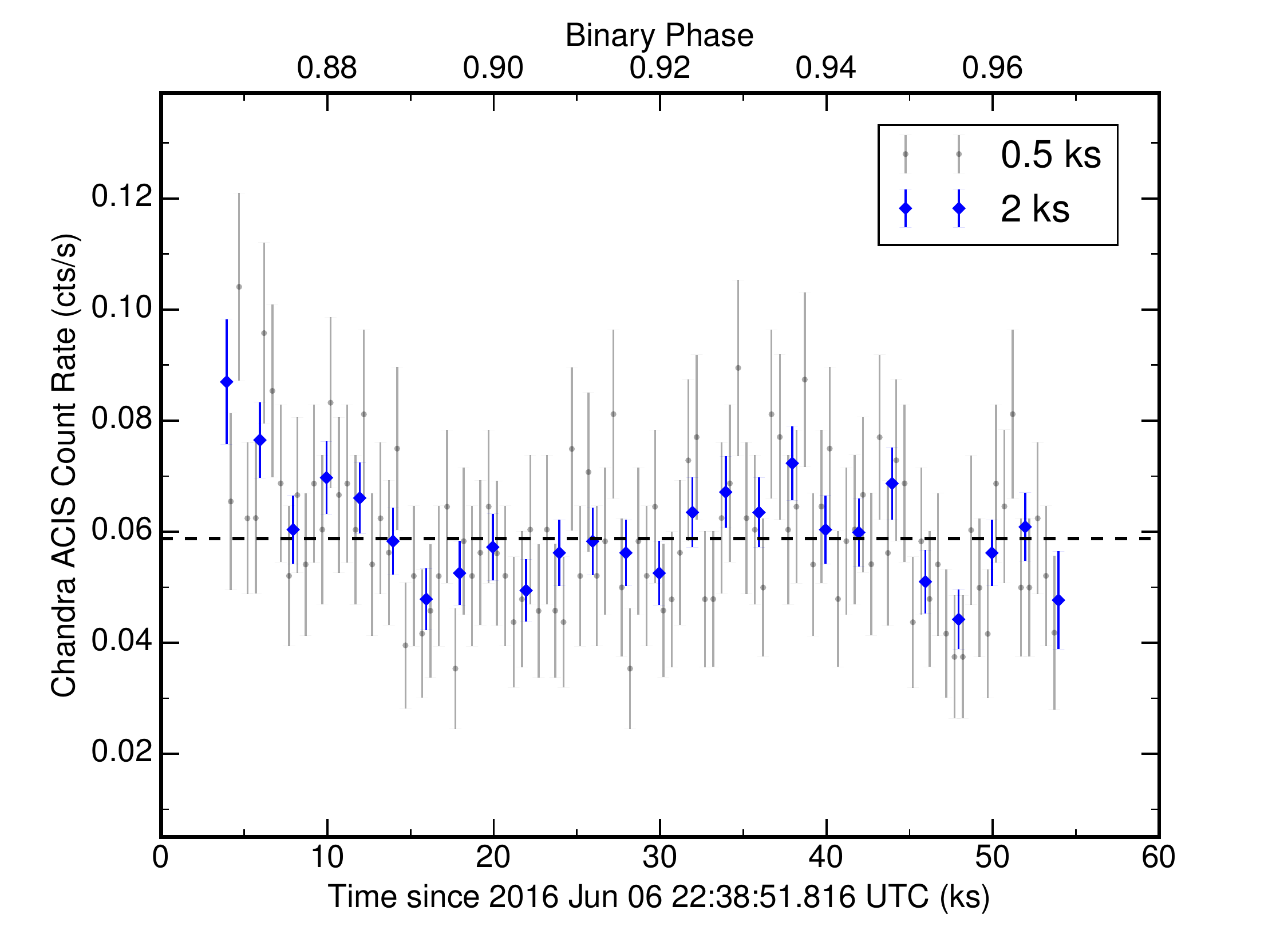}
    \caption{The Chandra ACIS light curve between 0.5--7 keV. The $\sim$50 ks exposure covers $\sim$11\% of the binary orbit. Photon arrival times have been barycenter corrected and are grouped into 500s (black) and 2ks (blue) bins. The black dashed line marks the weighted average of the 2ks dataset.}
    \label{fig:xray_lc}
\end{figure}

\subsubsection{X-ray Variability}
To analyze the light curve, we applied a barycenter correction to the observation times using the \texttt{CIAO} tool \texttt{axbary} and grouped the data into 0.5 and 2 ks time bins. The corresponding 0.5--7 keV light curves are shown in Figure~\ref{fig:xray_lc}.

Visual inspection of the light curves reveal the observed 0.5--7 keV count rates vary by a factor of two over this relatively small range of orbital phases. To quantify the significance of variability, we fit the 2 ks bins to a constant count rate model set to the weighted average of the dataset (Figure~\ref{fig:xray_lc}, dashed line). From a $\chi^2$ fit, we find a probability of $\sim$0.5\% that these data arise from a constant flux distribution. Data over a longer time period would be necessary to verify this variability and to determine whether the X-ray flux is modulated on the orbital period, as observed for many redbacks in the pulsar state, likely due to an intrabinary shock \citep{Bogdanov15}.

\subsubsection{Spectral Fitting}
\label{sec:spectralfitting}
We first attempted to fit our newly acquired \emph{Chandra} data with a purely thermal model, however, both an absorbed single blackbody (specified in \texttt{Xspec} as \texttt{tbabs*bbodyrad}) as well as a neutron star hydrogen atmosphere model (\texttt{tbabs*nsatmos}) yielded unsatisfactory fits ($\chi^2$/dof = 271.9/76 and 814.1/76, respectively).

Next, we fit the spectrum with an absorbed single power-law model (\texttt{tbabs*pegpwrlw}), finding an excellent fit ($\chi^2$/dof = 63.7/76, for a null hypothesis probability of 0.84 in \texttt{Xspec}) with photon index $\Gamma$=1.41~$\pm$~0.09 (90\% confidence). This value is consistent with the 2011 data, but with higher precision. In fitting for the absorption component, we found a neutral hydrogen column density ($N_{H}$) of (1.5 $\pm$ 0.6) $\times$ 10$^{21}$ cm$^{\textrm{-2}}$, which is about twice the Galactic column density in this direction inferred from $\textrm{\ion{H}{1}}$ data \citep[6.54$\times10^{20}\;\textrm{cm}^{\textrm{-2}}$,][]{Kalberla05}, and is consistent with the findings of \citet{Camilo16}, who inferred $N_{H}$ = 2.2$_{-1.3}^{+1.7}$ $\times$ 10$^{21}$ cm$^{\textrm{-2}}$ from an analysis of 2015 \emph{Swift} data (see \S\ref{sec:swiftdata}). Holding the absorption component fixed to the Galactic value yields a slightly worse fit ($\chi^2$/dof = 69.5/77). The extra absorbing material could be Galactic or could be intrinsic to the system, a possibility we discuss below. Given that the absolute value of $N_H$ is low and not all that well-constrained, for our spectral analysis we only quote fits in which $N_{H}$ was free to vary. The unabsorbed 1--10 keV flux of this model is $8.75\times10^{-13}$ erg cm$^{\textrm{-2}}$ s$^{\textrm{-1}}$, consistent with the 2011 \emph{Chandra} observations, which had large ($\gtrsim$30\%) uncertainties \citep{Strader15}.

While a simple power-law provides an acceptable fit to the data, we test for the presence of a thermal component in the spectra by adding a blackbody component (\texttt{tbabs*(pegpwrlw + bbodyrad)}). This model provides a slight improvement in the fit over the pure power-law model, with $\chi^2$/dof = 57.1/74, for a null hypothesis probability of 0.93. The power-law component of this model has a photon index $\Gamma=1.08^{+0.25}_{-0.15}$ while the blackbody has $kT = 0.60^{+0.15}_{-0.09}$. An $F$ test gives a probability of 1.7\% that the improvement in the fit that includes a thermal component occurs by chance if the pure power-law model was correct. Although this is not strong evidence for a thermal component in the spectrum, it is consistent with previous X-ray studies of tMSPs in their pulsar states \citep[e.g.,][]{Archibald10, Bogdanov11}, which also show marginal evidence for a thermal component coming from a small ``hot spot'' on the surface of the neutron star, likely from the heated magnetic polar caps. We also note the spectral index in this model is harder than those typically observed in tMSPs during their accretion states \citep[$\Gamma\sim$~1.7; e.g.,][]{deMartino10, Bogdanov15}, and is broadly consistent with synchrotron emission from an intrabinary shock.

\begin{figure}[t]
\centering
\subfloat{\includegraphics[angle=0,width=0.47\textwidth,
trim={0 0.0cm 0 0},clip]{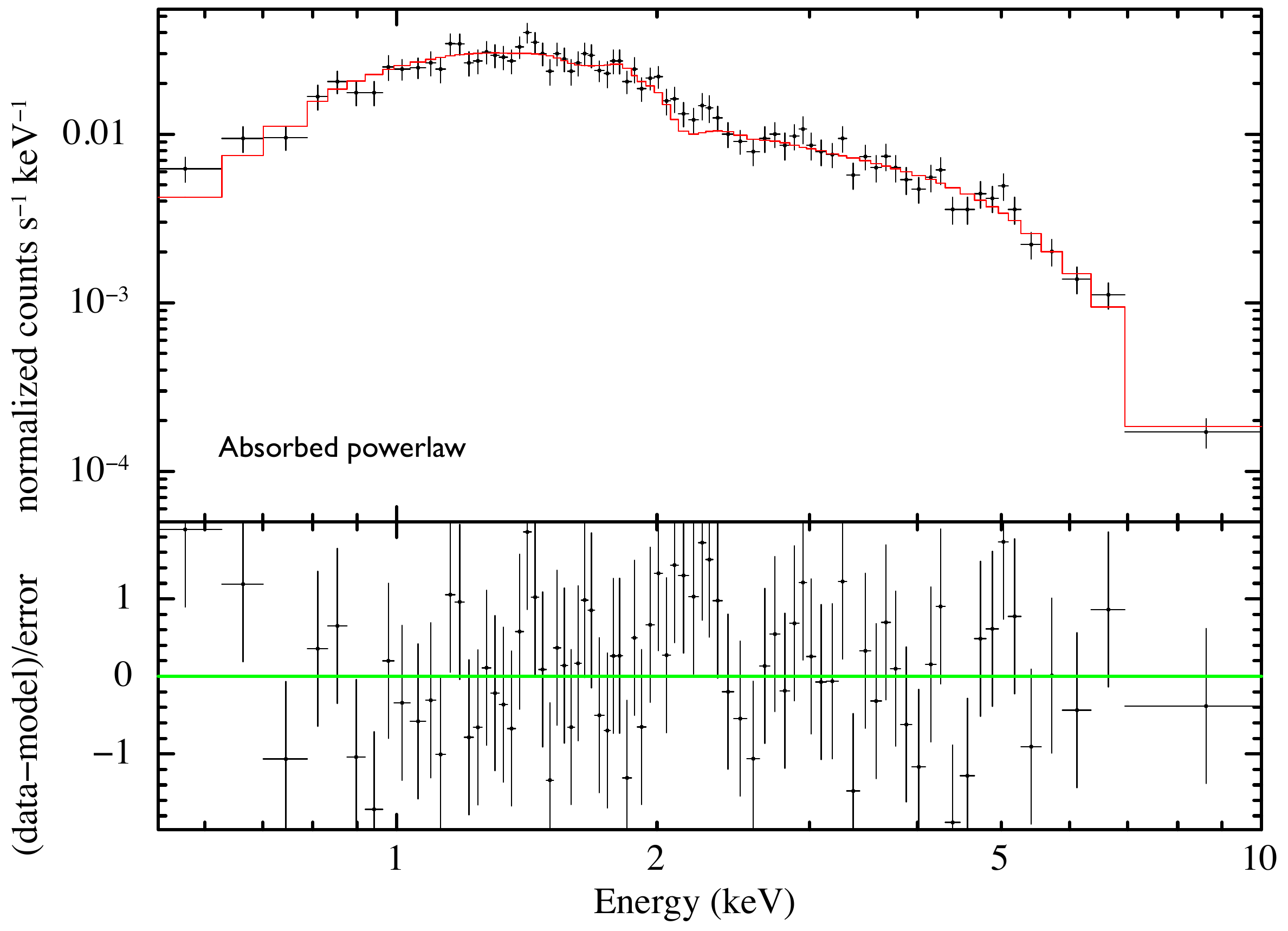}}\\
\subfloat{\includegraphics[angle=0,width=0.47\textwidth,
trim={0 0.0cm 0 0},clip]{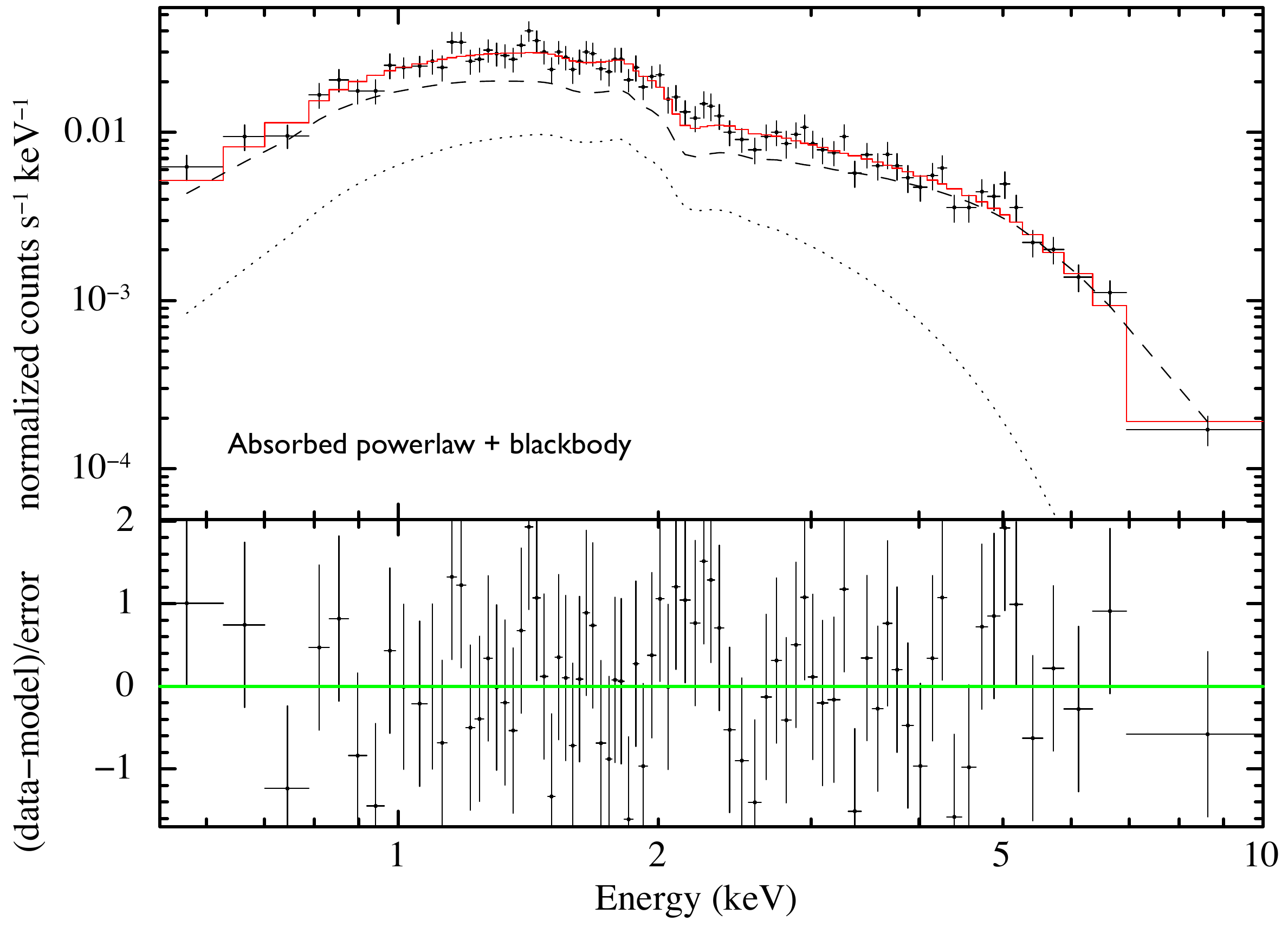}}

\caption{\emph{Chandra} X-ray spectra of J1417 with best fit models in red. In the upper figure, the data is fit with a pure absorbed power-law model, while the lower figure shows an absorbed power-law (dashed) plus the addition of a blackbody component (dotted). Fit residuals are shown in the bottom panels.}
\label{fig:chandra_specs}
\vspace{10pt}
\end{figure}

Both of these models are shown in Figure~\ref{fig:chandra_specs} and summarized in Table~\ref{table:xspecsummary}. All errors are quoted at 90\% confidence.

Assuming our new optical distance of 3.1 kpc (see \S\ref{sec:distance}), the unabsorbed flux of the pure power-law model corresponds to a 1--10 keV X-ray luminosity of (1.0$_{-0.04}^{+0.06}$)\,$\times$\,10$^{33}$ (d/3.1 kpc)$^{2}$ erg s$\mathrm{^{-1}}$. At the geometric distance derived from the \emph{Gaia} parallax (3.5 kpc, \S\ref{sec:distance}), the X-ray luminosity is $L_\textrm{x}$$\sim$1.3\,$\times$\,10$^{33}$ erg s$\mathrm{^{-1}}$. At either distance, J1417 is brighter in X-rays than all known redbacks in the pulsar state, likely due to a strong intrabinary shock. We return to this in \S\ref{sec:Halpha} and \S\ref{sec:discussion}.

\begin{deluxetable*}{lcc}[t]
\tablewidth{290pt}

\tablecaption{Summary of Chandra Spectral fits for J1417}
\tablehead{Parameters & power-law & power-law +\\
& & Thermal Component}

\startdata
$N_{H}$ (10$^{\textrm{21}}$ cm$^{\textrm{-2}}$) & 1.5 $\pm$ 0.6 & $<$1.0\\[3pt]
Photon Index ($\Gamma$) & 1.41 $\pm$ 0.09 & 1.08$_{-0.15}^{+0.25}$\\[3pt]
$F_{X}$ (10$^{-13}$ erg cm$^{\textrm{-2}}$ s$^{\textrm{-1}}$)\tablenotemark{a} & 8.75 $\pm$ 0.44 & 8.66 $\pm$ 0.50\\[3pt]
$kT$ (keV) & \nodata & 0.60$^{+0.15}_{-0.09}$\\[3pt]
Blackbody Fraction & \nodata & 0.13$^{+0.05}_{-0.09}$\\[3pt]
Blackbody Emission Radius (km)\tablenotemark{b} & \nodata & 0.09 $\pm$ 0.06\\[3pt]
$\chi^2_\nu$ (dof) & 0.84 (76) & 0.77 (74)

\enddata
\tablenotetext{a}{Unabsorbed X-ray flux between 1--10 keV}
\tablenotetext{b}{Assuming a distance of 3.1 kpc (\S\ref{sec:distance})}
\label{table:xspecsummary}
\end{deluxetable*}

\begin{deluxetable*}{lccr}[!h]
\tablecaption{\emph{Swift} XRT Observations of J1417}
\tablehead{
\colhead{Date} & \colhead{Exposure Time} & \colhead{XRT Count Rate} & Phase \\[3pt]
\colhead{(UT)} & \colhead{(ks)} & \colhead{(0.3--10 keV) (10$^{-2}$ s$^{-1}$)}} & 
\startdata
2016 May 07 & 1.14 & $2.0\pm0.5$ & 0.152 \\[3pt]
2016 Jun 05 & 0.94 & $1.7_{-0.5}^{+0.6}$ & 0.564 \\[3pt]
2016 Jul 05 & 1.05 & $1.9\pm0.5$ & 0.153 \\[3pt]
2016 Aug 02 & 0.88 & $2.7\pm0.6$ & 0.297 \\[3pt]
2016 Aug 31 & 0.95 & $1.5_{-0.4}^{+0.5}$ & 0.751 \\[3pt]
2016 Sep 29 & 0.98 & $1.6_{-0.4}^{+0.5}$ & 0.218

\enddata
\tablecomments{Uncertainties are listed at 90\% confidence. Orbital phase convention is that $\phi=0$ is the ascending node of the pulsar.}
\label{table:swiftobs}
\end{deluxetable*}

\subsection{Swift XRT}
\label{sec:swiftdata}
\citet{Camilo16} analyzed nine \emph{Swift} X-ray Telescope \citep[XRT,][]{Burrows05} observations of J1417 totalling 12.5 ks between 2015 Mar 13 and Jun 18. In addition to finding slight evidence for orbital variability in the light curve, the X-ray flux and spectral index they find are consistent with our new, deep \emph{Chandra} data.

We also observed this source with \emph{Swift}/XRT in photon counting mode on six epochs between 2016 May 7 and Sept 29 for a total of 5.9 ks. Similar to the analysis presented by \citet{Camilo16}, we extracted source counts in the 0.3--10 keV energy range from a circular region with radius 47$\arcsec$, with the background determined from a nearby circular, source-free region with radius 200$\arcsec$. We summarize these observations in Table~\ref{table:swiftobs}.

\begin{figure}[!b]
    \centering
    \includegraphics[width=1.0\linewidth,trim={0 0 0 0},clip]{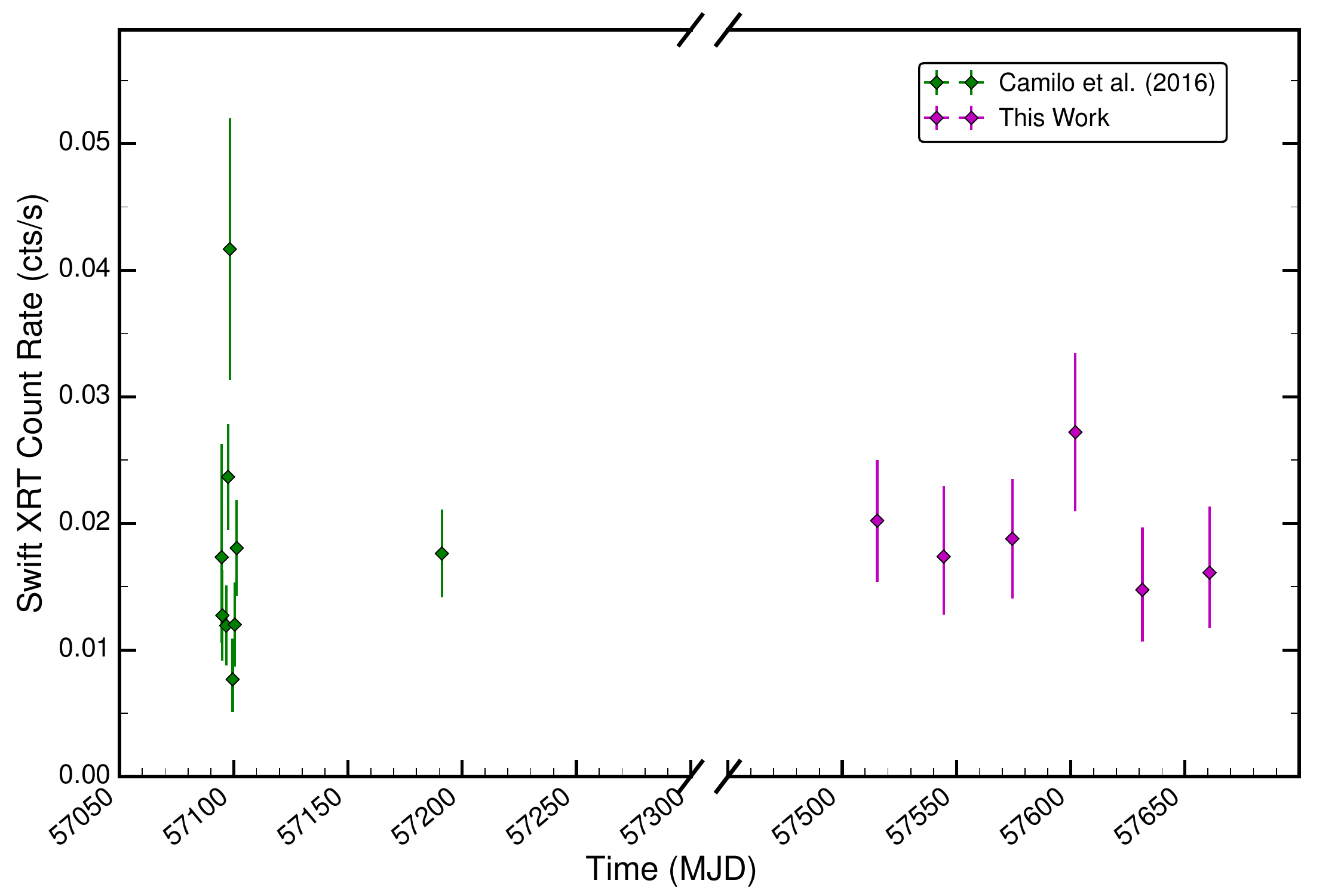}
    \caption{The 0.3--10 keV \emph{Swift} XRT count rates, extracted from 15 observations between 2015 Mar and 2016 Sept.}
    \label{fig:swiftlongLC}
\end{figure}

\begin{figure}[!t]
    \centering
    \includegraphics[width=1.0\linewidth,trim={0 0 0 0},clip]{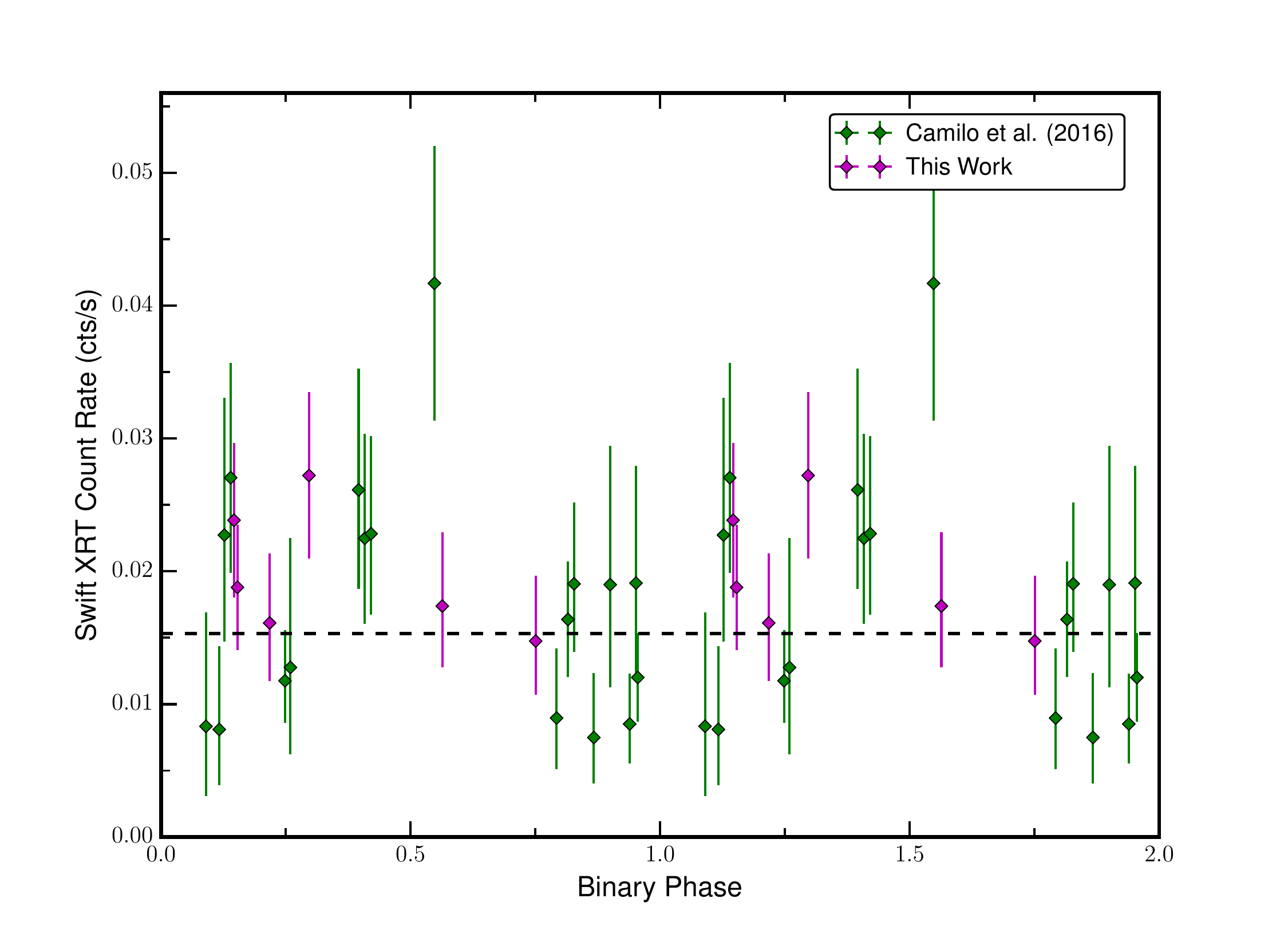}
    \caption{Snapshots from the 15 \emph{Swift} XRT observations in Figure~\ref{fig:swiftlongLC} folded on the orbital period of the binary. The times spanned by each exposure are smaller than the symbol sizes. The dashed line shows the weighted average of all snapshots. Two orbital cycles are shown for clarity.}
    \label{fig:swiftphasedLC}
    \vspace{10pt}
\end{figure}

Overall, we see that the mean 0.3--10 keV count rate of J1417 has remained relatively constant since the 2015 data were taken (Figure~\ref{fig:swiftlongLC}). These data also allow us to investigate the possibility of orbital variability. This is challenging for J1417, given its long period and the visibility constraints imposed by \emph{Swift's} low Earth orbit. We chose to analyze the folded X-ray light curve per individual ``snapshot'' so that each data point corresponds to a relatively narrow range of binary phases. The accuracy on the binary period derived from radio pulsar timing is 0.00003 d \citep{Camilo16}, such that the absolute phase uncertainty of the XRT observations is only $\sim$0.0005 (about 4 minutes) after building a phase-coherent solution forward to the most recent XRT epochs. This is negligible compared to the exposure times of each XRT snapshot. We show the folded XRT light curve in Figure~\ref{fig:swiftphasedLC}.

Before we discuss the phased light curve, it is worth a brief review of theoretical expectations for the orbital variations in the X-ray flux associated with an intrabinary shock. In the relatively simple model of \citet{Romani16}, the shock is assumed to occur between the (irradiation-driven) wind from the companion and the pulsar wind (see also the more sophisticated model of \citet{Wadiasingh17}). The controlling physical parameter is the ratio between the companion and pulsar wind momentum fluxes. If the pulsar wind dominates, the X-ray light curve has a double-peaked maximum around $\phi=0.25$ (when the companion is located between the MSP and the observer) and a minimum at $\phi=0.75$. If the companion dominates, the minimum and maximum are shifted by a phase of 0.5. The velocity of the companion wind with respect to the orbital velocity affects the symmetry of the X-ray light curves, and more face-on orbits mute the orbital variations.

Overall, the morphology of the J1417 X-ray light curve shows marginal evidence for features expected to arise due to an intrabinary shock, such as the ``peaks'' around $\phi \sim$ 0.15 and 0.35. If real, the double-peaked emission in this model is due to Doppler boosting along the line of sight, with a central dip due to the companion eclipsing part of the shock. However, we emphasize that the low count rate and incomplete phase coverage precludes a detailed comparison of the X-ray light curve with possible models. For example, the simple intrabinary shock model does not consider the possibility that the magnetic field of the secondary may be energetically relevant for shaping the interaction between the winds \citep[e.g.,][]{Roberts14, Sanchez17}.

\section{Radio Continuum Observations}
\label{sec:radioobs}
In all the known tMSPs that have been observed in their disk states, bright, flat-spectrum radio emission is present. This has been explained by partially self-absorbed synchrotron radiation coming from a compact jet-like outflow \citep[e.g.,][]{Hill11, Deller15, Papitto15}, though the short timescale anti-correlation between X-ray and radio emission in the tMSP J1023+0038 in its disk state clouds a simple jet interpretation for the radio emission \citep{Bogdanov17}.

In a number of black hole LMXBs, the connection between outflow processes and the inflow from an accretion disk have resulted in an empirical radio-X-ray luminosity ($L_{R}-L_{X}$) correlation, which ties the X-ray luminosity (a probe of the inner regions of the accretion flow) to the radio luminosity, which is set by the power of the jet \citep{Corbel00, Gallo03, Corbel13, Gallo14, Plotkin17, Plotkin17B}. A similar disk-jet coupling has also been observed (to a lesser extent) in a number of ``normal'' accreting neutron star systems, albeit with radio emission that is at least a factor of 10 dimmer than observed for black holes at the same X-ray luminosity \citep{Migliari03, Migliari06, Tudose09, Tetarenko16}. However, this similarity is obscured by the fact that different neutron star subsamples have shown different correlations in the $L_{R}-L_{X}$ plane, suggesting there may not be a ``universal'' $L_{R}-L_{X}$ correlation for neutron stars as there is for black holes \citep{Tudor17, Gallo18}.

Due to the less luminous jets in ``ordinary'' neutron star LMXBs compared to black holes \citep[by a factor $\sim$22,][]{Gallo18}, it is difficult to detect these systems in the radio at the relatively low X-ray luminosities we see in J1417. The vast majority of neutron star jets that have been explored to date have X-ray luminosities $\gtrsim$ 10$^{36}$ erg s$^{\textrm{-1}}$. In fact, only upper limits on the 5-GHz radio luminosity of a few neutron star LMXBs have been reported in the literature at X-ray luminosities $\lesssim$ 10$^{34}$ erg s$^{\textrm{-1}}$ \citep[e.g.,][]{Tudor17, Gallo18}. By contrast, while in their disk states, two of the three tMSPs have been firmly detected at 5-GHz with X-ray luminosities between 10$^{32}$--10$^{34}$ erg s$^{\textrm{-1}}$ \citep{Hill11, Deller15} \citep[the third tMSP is at much higher $L_X$,][]{Papitto13}. Despite the limited statistics, tMSPs appear to be more radio bright than typical neutron star LMXBs at the same X-ray luminosity, possibly due to the ejection of disk matter by a propeller mechanism \citep{Deller15}, though the existing radio upper limits for such systems at $L_X \sim$ 10$^{32}$--10$^{34}$ erg s$^{\textrm{-1}}$ are not particularly constraining. Furthermore, the anti-correlation between the radio/X-ray luminosity in the high and low X-ray modes of the accreting state of the tMSP J1023+0038 induces movement in the $L_{R}-L_{X}$ plane that is perpendicular to the slope of any mooted correlation \citep{Bogdanov17}, suggesting a crucial difference between tMSPs and ``normal'' accreting neutron star binaries.

\begin{deluxetable*}{lccccc}[!t]
\tablecaption{VLA radio continuum data for J1417}
\tablehead{
\colhead{Date} & \colhead{5.1 GHz flux density} & \colhead{7.1 GHz flux density} & spec. index & phase\\
\colhead{}	& \colhead{($\mu$Jy)} & \colhead{($\mu$Jy)} &  & }
\startdata

2015 Jun 15 & $<31.0$      & $<30.3$      & \nodata      & 0.269 \\[3pt]
2015 Jun 16 & $56.5\pm9.8$ & $40.9\pm9.5$ & $-1.0\pm0.9$ & 0.454 \\[3pt]
2015 Jun 17 & $<27.7$      & $<26.7$      & \nodata      & 0.640 \\[3pt]
2015 Jul 15 & $<29.9$      & $<31.4$      & \nodata      & 0.843 \\[3pt]
\hline
Combined\Tstrut    & $22.4\pm4.8$ & $13.5\pm5.1$ & $-1.5\pm1.3$ & \nodata

\enddata
\tablecomments{Radio upper limits are 3$\sigma$ values. Orbital phase convention is that $\phi=0$ is the ascending node of the pulsar.}
\label{table:VLAobs}
\end{deluxetable*}

\begin{deluxetable*}{lccccc}[!t]
\tablecaption{ATCA data for J1417}
\tablehead{
\colhead{Date} & \colhead{configuration} & \colhead{central freq.} & \colhead{flux density} & \colhead{spec. index}\\
\colhead{}	& \colhead{} & \colhead{(GHz)} & \colhead{($\mu$Jy)} & }
\startdata

 & & 1.43 & $294\pm42$ & \\[3pt]
2015 Oct 24 & 6A & 1.89 & $88\pm27$ & $-1.1\pm0.3$ \\[3pt]
 & & 2.59 & $145\pm21$ & \\[3pt]
\hline
 & & 1.41 & $258\pm90$ & \\[3pt]
\multirow{3}{*}{2016 Jun 06} & \multirow{3}{*}{1.5B} & 1.86 & $135\pm71$ & \multirow{3}{*}{$-1.2\pm0.8$} \\[3pt]
 & & 2.37 & $86\pm47$  & \\[3pt]
 & & 2.83 & $131\pm51$ &
\enddata
\label{table:ATCAobs}
\end{deluxetable*}



\subsection{VLA}
\label{sec:VLAdata}
Motivated by the evidence that J1417 was in a disk state based on the persistent double-peaked H$\alpha$ emission seen by \citet{Strader15}, and prior to the publication of a pulsar detection, we obtained VLA data to search for the presence of radio continuum jets. We observed J1417 with the Karl G. Jansky Very Large Array (VLA) under program VLA/15A-491 (PI Strader). Observations were obtained over four epochs during 2015 June/July, while the VLA was moving from BnA configuration into A configuration (which has a maximum baseline of $B_{\textrm{max}} = 36.4$ km). Each epoch was one hour in duration, yielding 38 minutes on source. The observations were obtained with 4096 MHz of bandwidth covering the entire C band (4--8 GHz), divided into 32 spectral windows each of 128 MHz bandwidth and sampled with 64 frequency channels. 

The VLA data were calibrated using J1352-4412 for complex gains, and 3C147 for absolute gain and bandpass. Data were edited, reduced, and imaged using standard routines in AIPS \citep{Greisen03} and CASA \citep[version 4.5.0;][]{McMullin07}. To minimize artifacts from wide fractional bandwidths and to constrain the spectral index, we divided the data into two frequency chunks (4--6 GHz, and 6--8 GHz), and imaged each chunk separately. The central frequencies of these subbands are 5.1 and 7.1 GHz, with beams of $1.91\arcsec \times 0.32\arcsec$ and $1.40\arcsec \times 0.27\arcsec$, respectively. We also note that these observations are all taken at very low elevation ($\lesssim 12^{\circ}$), as J1417 is near the southern declination limit of the VLA; these conditions somewhat hamper our sensitivity.

J1417 is detected in the combined 4 hr image at 5.1 GHz (22.4 $\pm$4.8 $\mu$Jy) and shows marginal evidence of radio emission at 7.1 GHz (13.5 $\pm$ 5.1 $\mu$Jy). However, after imaging the individual 1-hr blocks separately, it is clear that this combined flux density does not accurately represent the system: the source is not detected at either 5.1 or 7.1 GHz in three of the four blocks, while it is detected at high significance in both subbands during the second block (on UT 2015 Jun 16), with flux densities of 56.5 $\pm$ 9.8 (5.1 GHz) and 40.9 $\pm$ 9.5 $\mu$Jy, respectively (Table~\ref{table:VLAobs}). For flux density $S_{\nu} \propto \nu^{\alpha}$ with frequency $\nu$, we find a spectral index $\alpha$ = --1.0 $\pm$ 0.9 at this epoch. This spectral index does not constrain the nature of the source: it is consistent with either the steep-spectrum slope of an MSP \citep[$\alpha$ $\sim$ -1.6, e.g.,][]{Lorimer95,Maron00} or the flat spectrum emission observed for tMSPs in the disk state \citep{Papitto13, Bassa14, Archibald15, Deller15}. We summarize our VLA continuum observations in Table~\ref{table:VLAobs}.

We also separately imaged the individual $\sim$8 min scans of the block with the strongest 5.1 GHz detection (2015 Jun 16) to search for variability on these shorter timescales. The source was detected in some scans and not in others, consistent with its modest flux density and the higher rms noise of the individual scans. We found no evidence for statistically significant variations in flux density on timescales of minutes.

Although radio timing measurements do not always give an accurate flux density scale, \citet{Camilo16} report a 4.8 GHz flux density of $\sim 40$ $\mu$Jy for J1417 at one epoch of GBT observations, generally consistent with the flux density we observe during the epoch in which the source is detected. We discuss the observed variations in the flux density below in \S\ref{sec:discussion}.

\subsection{ATCA}
\label{sec:ATCAdata}
We observed J1417 with the Australia Telescope Compact Array (ATCA) on 2015 Oct 24, a few  months after our VLA observations, under program ID CX336 (PI Miller-Jones). We observed in the relatively extended 6A configuration ($B_{\textrm{max}}\sim 5.9$ km) with the 16cm receiver, providing coverage from 1--3 GHz. The bandwidth is sampled by 2048 frequency channels, each 1 MHz wide. We obtained 9.4 hours of data on-source, and calibrated the complex gains using PKS B1424--418 and the absolute flux and bandpass using PKS 1934--638. These data covered an orbital phase range of $\phi = 0.782$ to 0.824.

We also observed J1417 with ATCA on 2016 June 6 (program ID CX360, PI Miller-Jones), simultaneous with our \emph{Chandra} X-ray observations. The array was in the 1.5B configuration ($B_{\textrm{max}}\sim 4.3$ km), and we again observed with the 16cm receiver and the same correlator set-up and calibrators as the 2015 Oct observations. We obtained 5.0 hours of data on-source, over an orbital phase range of $\phi = 0.737$ to 0.760.

For both observations, the ATCA data were edited and reduced using standard routines in Miriad \citep{Sault95}. To minimize imaging artifacts caused by wide bandwidths, we split the data into frequency chunks (three for the 2015 Oct data and four for the 2016 June data). Each frequency chunk was self-calibrated and imaged in AIPS. We summarize our ATCA observations in Table~\ref{table:ATCAobs}.

For the 2015 Oct 24 data, the source was well-detected in each chunk, with flux densities of 294 $\pm$ 42 $\mu$Jy (1.43 GHz), 88 $\pm$ 27 $\mu$Jy (1.89 GHz), and 145 $\pm$ 21 $\mu$Jy (2.59 GHz). 
For a simple power-law fit these measurements imply a spectral index of --1.1 $\pm$ 0.3, in excellent agreement with the VLA measurement, though the flux densities together are not clearly well-described by this model. This may suggest the presence of frequency-dependent refractive scintillation \citep[e.g.,][]{Stinebring90}, frequency-dependent eclipses \citep[e.g.,][]{Broderick16}, or both.

For the 2016 June 6 data, the lower on-source time and more compact configuration led to a higher rms noise, and the source was only marginally detected at 2--3$\sigma$ in each chunk: 258 $\pm$ 90 $\mu$Jy (1.41 GHz), 135 $\pm$ 71 $\mu$Jy (1.86 GHz), 86 $\pm$ 47 $\mu$Jy (2.37 GHz), 131 $\pm$ 51$ \mu$Jy (2.83 GHz). Nevertheless, the spectral index implied by a power-law fit to these values is --1.2 $\pm$ 0.8, again consistent with the previous ATCA and VLA measurements. In fact, if we fit a single power-law with either the 2015 October or 2016 June ATCA data together with the 2015 Jun 16 VLA data, the best-fit spectral index is --1.0 $\pm$ 0.2 or --1.0 $\pm$ 0.3, broadly consistent with emission from a MSP and not from a LMXB jet.

\section{Optical Observations}
\label{sec:optobs}

\subsection{SOAR Spectroscopy}
We have continued occasional spectroscopic monitoring of J1417 with the Goodman spectrograph \citep{Clemens04} on the SOAR telescope, obtaining 10 epochs of spectroscopy from 2015 Jun 23 to 2017 Sep 01 (UT). The setup and reduction was identical to that from \citet{Strader15}, giving calibrated spectra with wavelength coverage of $\sim$ 5375--6640~\AA~at a resolution of about 1.7 \AA. We present a subset of these data and analyze them, along with a number of spectra obtained by \citet{Strader15} between 2014 Mar and 2015 Feb, in \S\ref{sec:Halpha}.

\subsection{Optical/Near-IR Photometry}
We obtained optical $BVI$ and near-IR $H$ photometry of J1417 using ANDICAM on the SMARTS 1.3-m telescope at CTIO roughly every night between 25 Jul and 19 Sep 2017 (UT). On each night, we took 200, 90, and 50 sec exposures in $B$, $V$, and $I$ bands, respectively. We simultaneously obtained dithered near-IR observations in $H$. After excluding poor frames, the on-source exposure time in $H$ was typically 120-200 sec per visit. Data were reduced following the procedures in \citet{Walter12}.

We performed differential aperture photometry to obtain instrumental magnitudes of the target source using fifteen ($BVI$) or seven ($H$) comparison stars as a reference. Absolute calibration was done with respect to the \citet{Landolt92} standard field TPheD ($BVI$) or to the 2MASS catalog ($H$). Our final dataset includes 36 measurements each in $BVI$ and 28 in $H$, with mean observed magnitudes $B = 17.18$, $V = 16.16$, $I = 15.00$, and $H = 13.38$. Median errors in each band are $\sim$~0.03 in $BVI$ and $\sim$~0.07 in $H$. Our full sample of SMARTS photometry is listed in Table~\ref{tab:SMARTSdata}. Observation epochs have been corrected to Barycentric Julian Date (BJD) on the Barycentric Dynamical Time (TDB) system \citep{Eastman10}.

\citet{Strader15}  obtained and analyzed time series photometry of J1417 in $BVR$ taken in mid-2014 using the PROMPT-5 telescope \citep{Reichart05}. This dataset included 29, 49, and 40 photometric measurements in $B$, $V$, and $R$ bands respectively. After applying a small zeropoint offset to PROMPT $B$ and $V$ to match the ANDICAM filters, we reanalyze this sample in conjunction with the SMARTS data in the following section.

\begin{deluxetable}{cccc}[!h]
\tablewidth{170pt}
\tablecaption{SMARTS Photometry of J1417}
\tablehead{BJD--2450000 & Band & mag & err \\
                   (d) & & & }

\startdata
7960.480563 & B & 17.143 & 0.022 \\
7960.483566 & V & 16.105 & 0.039 \\
7960.485358 & I & 14.979 & 0.056 \\
7960.481234 & H & 13.382 & 0.073 \\
7964.485152 & B & 17.167 & 0.149 \\
7964.488149 & V & 16.196 & 0.113 \\
7964.489947 & I & 15.045 & 0.080 \\
7964.485814 & H & 13.502 & 0.111 \\
7968.496390 & B & 17.190 & 0.039 \\
... & ... & ... & ... \\
\enddata
\tablecomments{The full SMARTS dataset is available in machine-readable format. We show a portion of the table here as a preview of its form and content. Magnitudes have not been corrected for extinction.}
\label{tab:SMARTSdata}
\end{deluxetable}

\begin{figure*}[t]
    \centering
    \includegraphics[width=1\textwidth,trim={0 0 0 0},clip]{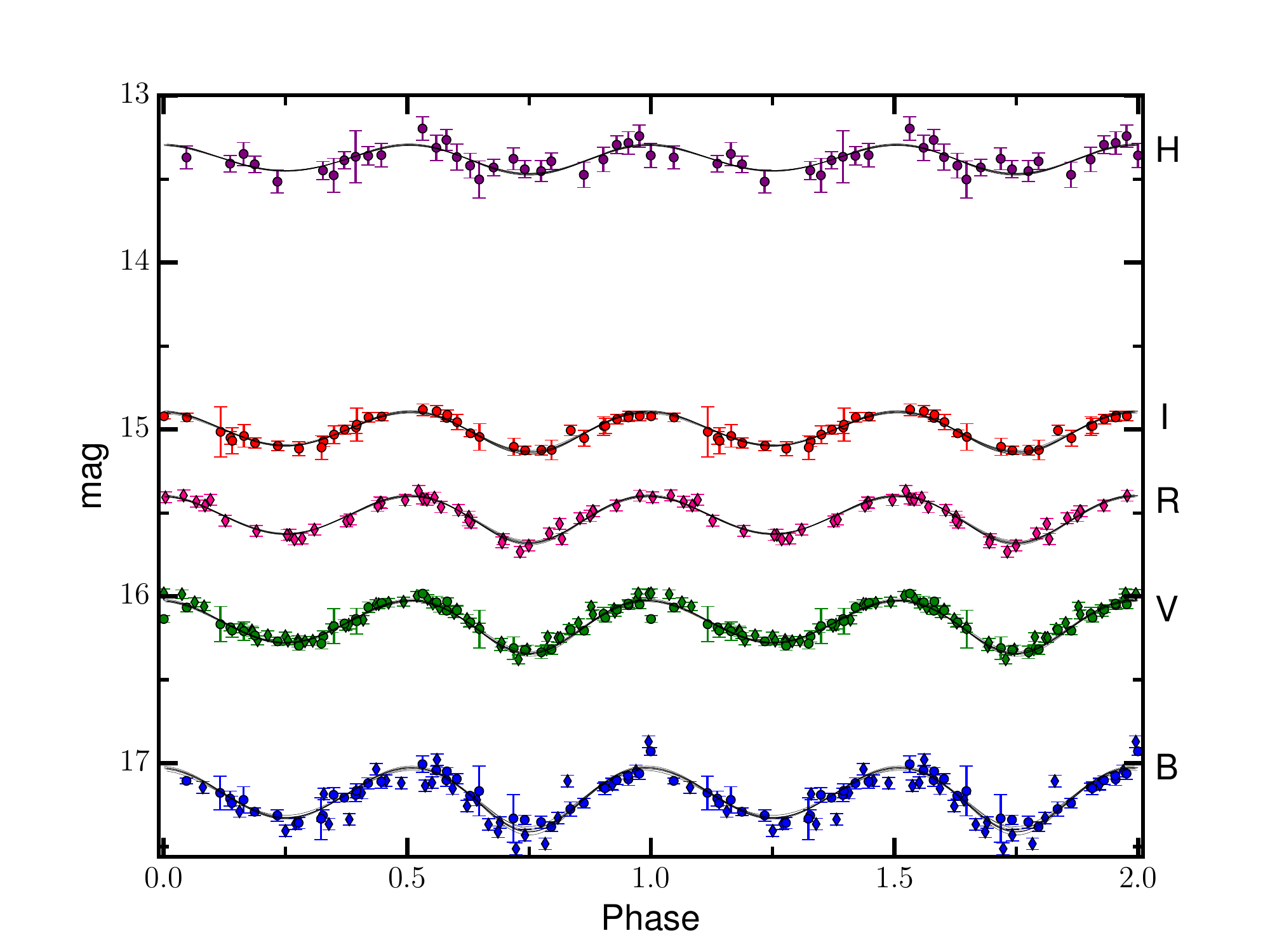}
    \caption{PROMPT (diamonds) and SMARTS (circles) optical/near-IR photometry with best fit ELC models. Two orbital phases are shown for clarity. The black lines show 8 random samples from the posterior marginalized over all the free parameters in the SMARTS+PROMPT ``No Disk'' MCMC run (Table \ref{table:lcmodels}, top panel, column 1).}
    \label{fig:lcmodel}
    \vspace{10pt}
\end{figure*}

\section{Light Curve Fitting}
\label{sec:optLCs}
\citet{Strader15} analyzed the PROMPT $BVR$ photometry under the assumption that the light curves were dominated by ellipsoidal variations due to the tidal deformation of a Roche lobe filling secondary, finding that the inclination was $i$ = 58~$\pm$~2$^{\circ}$. Here we relax these assumptions and provide more comprehensive light curve fits to the full photometric dataset discussed above. 

Although the overall scale of the binary system was already well constrained by the spectroscopic results presented in \citet{Strader15}, in our light curve fits we adopt the pulsar ephemeris from \citet{Camilo16}, which provide improved precision on the orbital period and mass ratio. Throughout our analysis, we assume a circular orbit (consistent with the optical radial velocity curve), and fix the orbital period $P_b$, radial velocity semi-amplitude of the secondary $K_2$, mass ratio $q$, and projected semimajor axis of the pulsar orbit $x_{1}$, to the values presented by \citet{Strader15} or \citet{Camilo16}: $P_b$ = 5.37372 d, $K_2$ = 115.7 km s$^{\textrm{-1}}$, $q$ = $M_{2}/M_{1}$ = 0.171, and $x_{1} = a_{1}\,\textrm{sin}\,i$ = 4.876(9) lt-s. 

We model the light curves with the Eclipsing Light Curve \citep[ELC;][]{ELCcode} code, which fits normalized, filter-specific light curves in each band independently. We fit two main classes of models. The first assumes no accretion disk is present. In these ``No Disk'' fits, the free parameters are the inclination $i$, the Roche lobe filling factor of the secondary $f_2$, and its intensity-weighted mean temperature $T_2$. Given the possible evidence for a disk from the H$\alpha$ spectroscopy, we also explore fits in which a non-negligible accretion disk component was present. Under these circumstances, the secondary was assumed to completely fill its Roche lobe ($f_2 = 1$), while the disk is characterized by five parameters: the inner disk temperature $T_{\textrm{disk}}$, the inner ($r_{\textrm{in}}$) and outer ($r_{\textrm{out}}$) radii of the disk, the power-law index of the disk temperature profile $\xi$, and the opening angle of the disk rim $\beta$.

We present fits using both the SMARTS data only and with both datasets together; given the consistency of the two datasets, using both together provides the best time and phase coverage. These ELC fits were done using its included Markov Chain Monte Carlo (MCMC) sampler, and we summarize the best-fit values as the median of the posterior distribution, with $1\sigma$ equivalent uncertainties given as the 15.9 and 84.1\% quantiles of the posterior. These values are given in Table~\ref{table:lcmodels}, and sample models are plotted in Figure~\ref{fig:lcmodel}.

Following the recommendations of \citet{Hogg17} for reporting results of an MCMC run, we refrain from showing only a ``best-fit'' model, taken as the collective medians of each individual posterior distribution. Instead, we have selected a few randomly drawn example samples from the posterior of our MCMC run marginalized over all parameters, and plot these with the data in Figure~\ref{fig:lcmodel}. This ensures we retain probabilistic information about the sampling results while also showing models that are satisfactory fits to the data.

The first main result from the fits is that there is no evidence for a disk in the optical/near-IR light curves: its inclusion does not improve the quality of the model fits, and in the best-fitting case it makes up a tiny fraction of the light ($\sim 1\%$ in $V$). Either a disk is not present or it is so faint compared to the giant secondary that its presence is irrelevant in the broadband photometry.

Considering the fits with no disk, the main result is the well-known degeneracy between the inclination and the filling factor. In the absence of significant irradiation, the relative amplitude of ellipsoidal modulations depends mainly on the observable projected area of the secondary. Therefore, light curve models with lower inclinations (more face-on) and larger filling factors will look essentially identical to models of more inclined systems whose secondary fills a smaller fraction of its Roche lobe.

We find a filling factor $f_2 = 0.83^{+0.05}_{-0.07}$ and an associated inclination 64\fdg2{$^{+5.0}_{-7.8}$}. The slightly higher (and more uncertain) inclination than found in \citet{Strader15} leads to a slightly lower primary mass: $M_1 = 1.62^{+0.43}_{-0.17} M_{\odot}$, but with a larger uncertainty. If instead we fix the secondary to fill its Roche lobe (Table~\ref{table:lcmodels}, top panel, column 2), then $i$ = 55\fdg2{$^{+0.7}_{-1.0}$} and $M_1 = 2.13^{+0.08}_{-0.06} M_{\odot}$, consistent with the neutron star mass found by \citet{Strader15}.

In addition to the above models, we also explored ones that included irradiation. This would appear in the optical/near-IR light curves as extra luminosity on the ``day'' side (the side of the tidally locked companion facing the pulsar) as the star is heated directly by high-energy photons from the pulsar \citep[e.g.,][]{Romani16}, or indirectly through reprocessing of the pulsar spin-down power in an intra-binary shock \citep[e.g.,][]{Romani15_2, Romani15, Rivera17, Wadiasingh17}.

It is evident from our light curves that their behavior is well-described by standard ellipsoidal variations, with no obvious evidence for irradiation. This was reflected in our ELC fits that included irradiation as a free parameter, which found a negligible contribution from an irradiating source. Obviously, high-energy emission is present in this system, but we find no sign that it affects the optical/near-IR light curves.

Overall, we have improved on the light curve fitting presented by \citet{Strader15} by exploring a much larger parameter space in conjunction with additional data and more precise dynamical constraints. The model with the fewest components that best represents the data is the SMARTS+PROMPT ``No Disk'' model with the filling factor left to vary (Table~\ref{table:lcmodels}, top panel, column 1), which we focus on for the remainder of the paper. The component masses associated with this fit are $M_{1} = 1.62_{-0.17}^{+0.43}\,M_{\odot}$ and $M_{2} = 0.28_{-0.03}^{+0.07}\,M_{\odot}$, well within the range of typical values inferred for other neutron star binaries associated with \emph{Fermi} sources \citep[e.g.,][]{Ransom11, Romani15, Strader16, Halpern17, Shahbaz17}.

\begin{deluxetable*}{@{\extracolsep{0pt}}c||l|cccc}[!t]
\tablewidth{490pt}

\tablecaption{Summary of ELC fits for J1417}

\startdata
\hline \hline
& & SMARTS+PROMPT & SMARTS+PROMPT & SMARTS+PROMPT & SMARTS Only \\[3pt]
& & & \scriptsize{($f_2=1.0$ fixed)} & \scriptsize{($f_2=0.83$ fixed)} & \\[3pt]
\cline{2-6}
 & Fitted Parameters\Tstrut & & & & \\[3pt]\cline{2-2}
 & incl ($^\circ$) & 64.2$_{-7.8}^{+5.0}$ & 55.2$_{-1.0}^{+0.7}$ & 64.0$_{-1.4}^{+1.1}$ & 62.3$_{-9.6}^{+6.7}$ \\[4pt]
 & R.L filling factor ($f_2$) & 0.83$_{-0.07}^{+0.05}$ & 1.0\tablenotemark{a} & 0.83\tablenotemark{a} & 0.83$_{-0.10}^{+0.06}$ \\[4pt]
 \multirow{4}{*}{\Large{No Disk}} & $T_2$ (K) & 4560$_{-336}^{+460}$ & 4523$_{-317}^{+297}$ & 4559$_{-344}^{+270}$ & 4509$_{-415}^{+388}$ \\[4pt]
 & $\chi_{\nu}^{2}$(dof) & 386.1(250) & 390.8(251) & 386.1(251) & 127.1(132) \\[4pt]\cline{2-2}
 & Derived Parameters\Tstrut & & & & \\[3pt]\cline{2-2}
 & $M_1$ (\Msun) & 1.62$_{-0.17}^{+0.43}$ & 2.13$_{-0.06}^{+0.08}$ & 1.63$_{-0.05}^{+0.06}$ & 1.70$_{-0.25}^{+0.65}$ \\[4pt]
 & $M_2$ (\Msun) & 0.28$_{-0.03}^{+0.07}$ & 0.36 $\pm$ 0.01 & 0.28 $\pm$ 0.01 & 0.29$_{-0.04}^{+0.11}$ \\[4pt]
 & $R_2$ ($R_{\odot}$) & 3.66$_{-0.33}^{+0.30}$ & 4.20$_{-0.04}^{+0.05}$ & 3.66$_{-0.03}^{+0.04}$ & 3.72$_{-0.64}^{+0.52}$ \\[4pt]
 & log(g) (cgs) & 2.76 $\pm$ 0.01 & 2.75 $\pm$ 0.01 & 2.75 $\pm$ 0.01 & 2.76 $\pm$ 0.01 \\[4pt]
 \hline \hline
 & Fitted Parameters\Tstrut & & & & \\[3pt]\cline{2-2}
 & incl ($^\circ$) & 56.0$_{-2.1}^{+1.4}$ & ... & ... & 54.7$_{-3.4}^{+2.2}$ \\[4pt]
 & $T_2$ (K) & 4633$_{-536}^{+259}$ & ... & ... & 4679$_{-684}^{+1118}$ \\[4pt]
 & $T_{\textrm{disk}}$ (K) & 3191$_{-2159}^{+1480}$ & ... & ... & 3261$_{-3095}^{+1919}$ \\[4pt]
 \multirow{8}{*}{\Large{Disk}} & $r_{\textrm{in}}$\tablenotemark{b} & 0.22$_{-0.18}^{+0.20}$ & ... & ... & 0.25$_{-0.18}^{+0.22}$ \\[4pt]
 & $r_{\textrm{out}}$\tablenotemark{b} & 0.58$_{-0.17}^{+0.20}$ & ... & ... & 0.66$_{-0.22}^{+0.18}$ \\[4pt]
 & $\beta$ ($^{\circ}$) & 15.1$_{-14.3}^{+5.7}$ & ... & ... & 14.6$_{-14.1}^{+4.5}$ \\[4pt]
 & $\xi$ & -0.75$_{-0.23}^{+0.18}$ & ... & ... & -0.74$_{-0.27}^{+0.39}$ \\[4pt]
 & $\chi_{\nu}^{2}$(dof) & 388.0(246) & ... & ... & 127.9(128) \\[4pt]\cline{2-2}
 & Derived Parameters\Tstrut & & & & \\[3pt]\cline{2-2}
 & $M_1$ (\Msun) & 2.08 $\pm$ 0.15 & ... & ... & 2.18$_{-0.29}^{+0.28}$ \\[4pt]
 & $M_2$ (\Msun) & 0.36 $\pm$ 0.03 & ... & ... & 0.37 $\pm$ 0.05 \\[4pt]
 & $R_2$ ($R_{\odot}$) & 4.16 $\pm$ 0.10 & ... & ... & 4.23$_{-0.20}^{+0.18}$ \\[4pt]
 & log(g) (cgs) & 2.75 $\pm$ 0.01 & ... & ... & 2.76 $\pm$ 0.02
\enddata
\tablenotetext{a}{Filling factor held constant}
\tablenotetext{b}{Inner and outer disk radii are expressed as a fraction of the effective Roche-lobe of the primary}
\label{table:lcmodels}
\end{deluxetable*}

\subsection{Distance}
\label{sec:distance}
Following similar steps as detailed in \citet{Strader15} and \citet{Swihart17}, we estimate the distance to J1417 by comparing its observed magnitude to its intrinsic luminosity derived from the best-fit radius and effective temperature of the secondary from our optical/near-IR light curve models. After applying bolometric corrections to each band as a function of temperature and metallicity from fitting 10 Gyr, [Fe/H] = --0.65 isochrones \citep{Marigo08}, we compared the extinction-corrected magnitudes (corrected using the \citet{S+F11} reddening maps) to the predicted absolute magnitude in each band. Given the inferred effective temperature and Roche lobe filling factor and their uncertainties (Table~\ref{table:lcmodels}, top panel, column 1), we find a distance of 3.1 $\pm$ 0.6 kpc. For a Roche lobe filling secondary ($f_2$=1.0) at the same temperature, the distance would instead be 3.6 $\pm$ 0.5 kpc.

From consideration of the mean colors and two low-resolution spectra of J1417, \citet{Strader15} assumed the effective temperature of the secondary is around 5000K, about 450 K warmer than the effective temperature we find in the new light curve analysis. This warmer temperature, combined with the assumption of a Roche Lobe-filling secondary, led to a larger, brighter star and a greater inferred distance of 4.4 kpc.

As mentioned above, these fits were performed at the subsolar metallicity inferred from the optical spectrum \citep{Strader15}, but we note that assuming solar metallicity instead produces consistent distances to within 0.1 kpc.

J1417 is present in \emph{Gaia} DR2\footnote{https://www.cosmos.esa.int/web/gaia/dr2} with parallax 0.22 $\pm$ 0.07 mas. Given the low significance of this measurement, we use a Bayesian inference procedure recommended by \citet{BailerJones18} to estimate the distance to J1417. Using the weak distance prior of \citet{Gandhi18} based on the expected distribution of LMXBs in the Galaxy, the geometric distance estimate is 4.2$_{-1.1}^{+2.2}$ kpc (1$\sigma$). \citet{Jennings18} adopt a distance prior based on a MSP-specific model for the space density, estimating 3.50$_{-0.38}^{+2.09}$ kpc. Both these values are fully consistent with our distance estimates from the light curve model, but not with the DM-derived distance (1.6 kpc) from \citet{Camilo16}. For the remainder of the paper, we adopt the distance inferred from our light curve fits (3.1 $\pm$ 0.6 kpc).

\section{H$\alpha$ Spectroscopy}
\label{sec:Halpha}

\subsection{The Spectra Themselves}

\begin{figure*}[]
    \centering
    \includegraphics[width=0.75\textwidth,trim={0 0 0 0},clip]{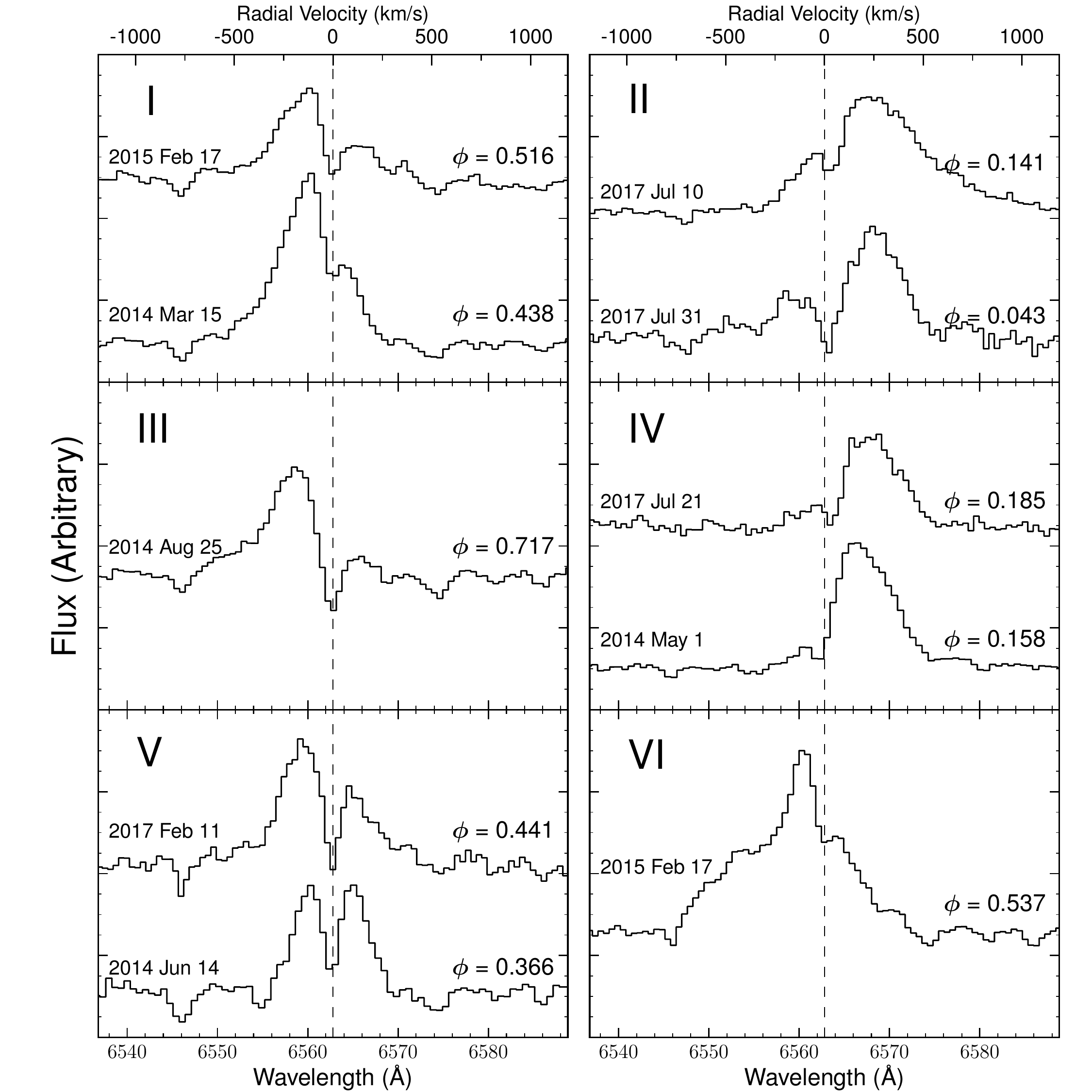}
    \caption{Example optical spectra of J1417 around the H$\alpha$ region. All spectra have been shifted to the rest frame of the secondary (dashed line). Each panel shows spectra that are representative of nearly all the spectra obtained at that orbital phase.}
    \label{fig:Halpha}
    \vspace{10pt}
\end{figure*}

\begin{figure*}[]
    \centering
    \includegraphics[width=0.75\textwidth,trim={0 0 0 0},clip]{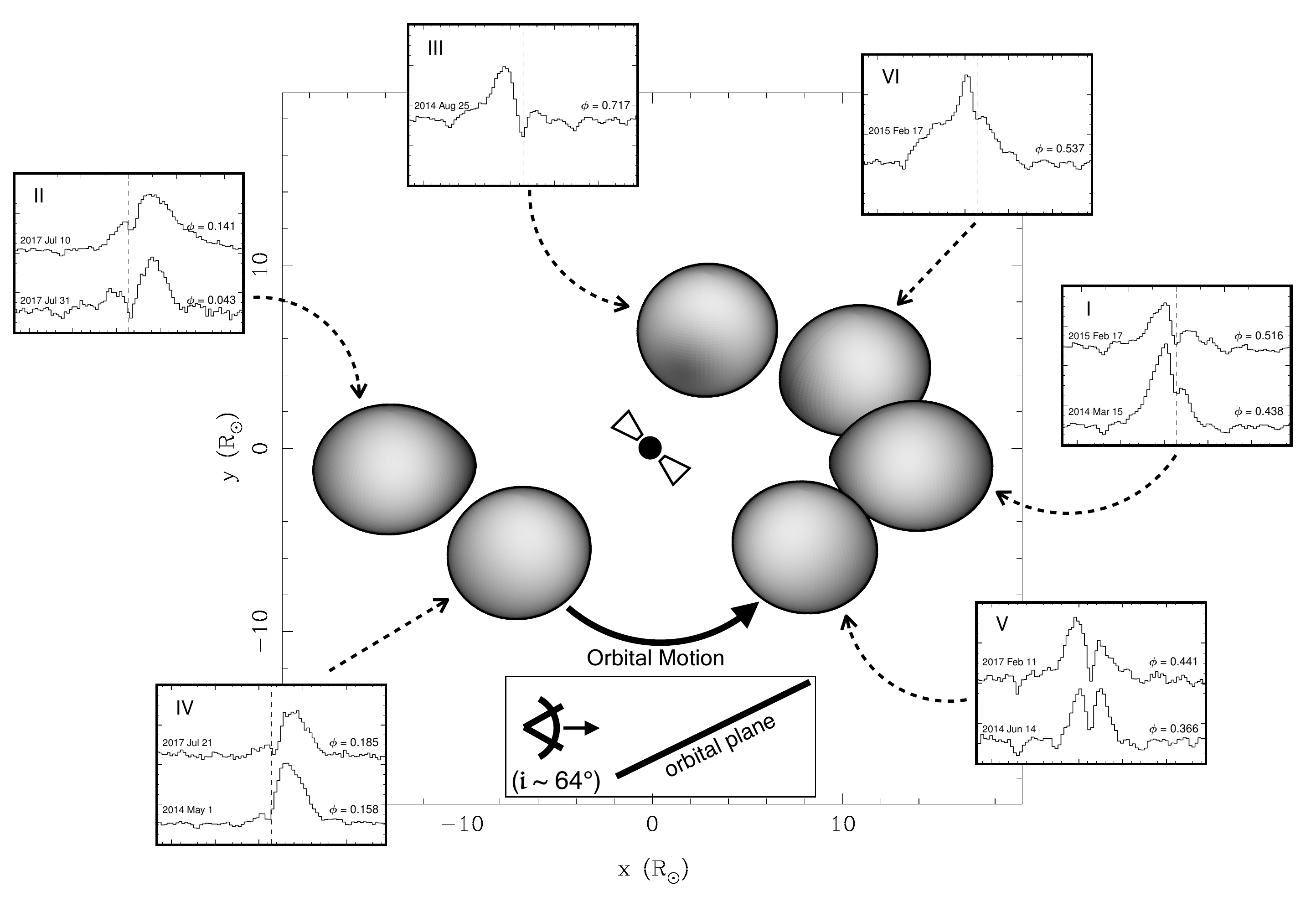}
    \caption{A schematic diagram of J1417 (as seen from Earth) showing the orbital position of the secondary corresponding to the range of H$\alpha$ phenomenology we show in the panels of Figure~\ref{fig:Halpha}.}
    \label{fig:drawing}
    \vspace{10pt}
\end{figure*}

In tMSPs, the presence of a broad, double-peaked H$\alpha$ emission profile has generally been accepted as evidence for an accretion disk \citep[e.g.,][]{Wang09, Bogdanov15}. Yet the observation of J1417 as a radio pulsar, and the other evidence that this system is in the pulsar state (i.e. the X-ray light curve and radio spectral index), is in opposition to this evidence for a disk. Here we carefully consider the morphology of the H$\alpha$ emission in J1417 and discuss models that can explain its origin. 

First, we echo the basic statement in \citet{Strader15} that the majority of the H$\alpha$ profiles show two peaks. However, we also find that the components vary in both shape and amplitude over time, and that these variations appear to be associated with the orbital phase of the binary in a consistent manner over the 3.5-year time baseline of our optical spectra.

In Figure~\ref{fig:Halpha} we show example spectra of J1417 around the H$\alpha$ region. These have all been shifted to the rest frame of the secondary. We show six panels, each with one or two spectra, that are representative 
of nearly all the spectra obtained at that phase. In panels I \& II, the emission is double-peaked but asymmetric, with stronger blue or red components, respectively. Here the phases are near 0.5 and 0 (at quadrature). In III \& IV (taken near conjunction) one of the components has almost completely vanished. In general, when the secondary is receding from Earth along our line-of-sight ($\phi$=0.25--0.75), the H$\alpha$ profile has a stronger blue component when compared to the red (Figure~\ref{fig:Halpha}, panels I, III), and vice versa when the companion approaches Earth ($\phi$=0.0--0.25 \& 0.75--1.0; panels II, IV).

Panel V shows two spectra taken between these extremes, which shows nearly symmetric H$\alpha$ emission. Panel VI shows the most outlying spectrum observed, with unique phenomenology in our sample: it has a bright, narrow main peak superimposed on a broader, asymmetric component. In Figure~\ref{fig:drawing}, we show a schematic diagram of the binary at the orbital phases corresponding to the six panels in Figure~\ref{fig:Halpha}.

For a typical H$\alpha$ equivalent width of about 5\AA\ and the range of distances from the last section, we find that the H$\alpha$ luminosity is $\sim 10^{31}$ erg s$^{-1}$.

\subsection{Understanding the H$\alpha$}

Now we consider possible origins of the H$\alpha$ emission and how it would vary with orbital phase. 
H$\alpha$ emission might conceivably originate (1) from the outer regions of an accretion disk, (2) from the chromosphere of the secondary, (3) in material lost from the secondary, such as an outflow or wind, possibly enhanced by an intrabinary shock between the pulsar and the secondary. We consider these in turn.

\subsubsection{A Disk}

Double-peaked Balmer emission is generally interpreted as arising from an accretion disk, and has been observed for the tMSPs PSR J1023+0023 and XSS J12270--4859 in their disk states \citep[e.g.,][]{Wang09, deMartino14}. Such emission is not observed during the pulsar state of these systems, presumably due to the disappearance of the disk. The spectra obtained of PSR J1023+0023 in its disk state in 2013 Oct \citep{Bogdanov15} show slightly asymmetric peaks, with a stronger blue component, and the overall shape and binary phase of these profiles are roughly consistent with the spectra we show in panels I \& V of Figure~\ref{fig:Halpha}.

\citet{Zelati14} also show the H$\alpha$ profile of PSR J1023+0038 in its disk state taken a few months after \citet{Halpern13}. The emission is still double-peaked, but with a slightly larger red peak than the blue, similar to Figure~\ref{fig:Halpha}, panel II. However, the spectrum they present is a co-addition of four exposures taken at a number of binary phases ($\phi$$\sim$0.25--0.6, Coti Zelati, private communication) so
we cannot make a direct phase-resolved comparison to J1417.

However, several pieces of evidence suggest that the J1417 H$\alpha$ emission is unlikely to originate from a standard LMXB accretion disk. First, the minimum of the double-peaked profile is stationary in the rest frame of the secondary. This might conceivably be explained if the secondary has strong H$\alpha$ absorption that sweeps through an essentially fixed emission profile (the primary motion is only $\sim 20$ km s$^{-1}$), but this does not explain why the emission components \emph{also} appear to lie at essentially the same velocity as a function of orbital phase in the secondary rest frame.

Second, as pointed out by \citet{Strader15}, the peak separation when the ``classic" double-peaked profile is evident (Figure~\ref{fig:Halpha}, panel V) is comparable to the orbital velocity, which would unphysically suggest that the disk fills the entire binary. (At other phases the peak separation is broader and more easily accommodated.) 

Finally, there is no independent evidence of a disk: one is not necessary to fit the optical/near-IR light curves, and \citet{Camilo16} find no evidence for UV emission from a disk using \emph{Swift}/UVOT observations.
We do note that these constraints are weaker than for typical redbacks owing to the higher luminosity of the giant.

\subsubsection{Stellar Chromosphere}

If the emission were coming directly from the atmosphere of the giant, as is the case in chromospherically active stars such as FK Com and RS CVn systems \citep{Drake06}, the H$\alpha$ emission would be relatively narrow and track the secondary's motion. It would not be expected to vary substantially with orbital phase. Instead, the H$\alpha$ emission we observe is broad and variable in velocity, and is never observed at the systemic velocity of the secondary, which instead shows (likely intrinsic) H$\alpha$ absorption. Thus the chromosphere of the secondary is not the main contributor to the emission. 

\subsubsection{Stellar Wind}

An outflow or wind from the secondary would form a natural medium for the production of an H$\alpha$ profile that varies, but is centered on the velocity of the secondary. The observed rotational velocity of the star is consistent with the star being tidally locked to the neutron star \citep{Strader15}, as expected for a close compact binary in which extensive mass transfer has occurred. For low-mass stars like J1417, this rapid rotation can result in a strong dynamo that can boost the surface magnetic fields by a factor of $\sim 10^2$--$10^3$ compared to similar isolated stars \citep{Morin12}. This enhanced surface field 
can lead to strong magnetically driven winds, since mass loss from low-mass red giants is caused primarily by 
Alfv\'{e}n wave pressure \citep{Cranmer11}.

For J1417 specifically, using the values in Table~\ref{table:lcmodels}, the escape velocity from the secondary is no larger than 170--180 km s$^{-1}$. For all of the emission profiles a substantial (and sometimes dominant) amount of the emission is present at velocities larger than the escape velocity of the secondary, showing that the gas is not simply an outflow but a wind that has escaped the red giant.

The wind from J1417 could emit H$\alpha$ photons directly given that a source of ionization is present, or indirectly
as it interacts with the pulsar wind, forming an intrabinary shock near the companion surface \citep{Romani16,Wadiasingh17}.
Such shocks are likely responsible for at least some of the observed X-ray variability in redbacks and the known tMSPs \citep[e.g.,][]{Roberts14} and provide a natural location for a region of gas with high enough temperature to emit H$\alpha$ photons.

This latter scenario can at least partially explain the complicated morphology of the emission: the red giant wind would not flow away from the star unimpeded, but instead would be halted and swept away by the pulsar wind, perhaps wrapping around the binary and extending out beyond the orbit of the secondary.

This ``radio ejection'' mechanism \citep{Burderi02} has been used successfully to  explain the H$\alpha$ emission in the globular cluster binary PSR J1740-5340A, which hosts a stripped subgiant in a 1.3 day orbit \citep{D'Amico01, Bogdanov10, Mucciarelli13}.
\citet{Sabbi03} analyzed the phase-resolved H$\alpha$ emission profiles of the companion to PSR J1740-5340A, finding that a combination of material flowing to the inner Lagrange point and gas swept out beyond the companion's orbit could explain the broad components not associated with the stellar atmosphere. The overall shape and binary phase of some of our spectra are remarkably similar to the profiles seen in the optical spectra of PSR J1740-5340A. In particular, the spectrum we present in panel VI of Figure~\ref{fig:Halpha} shows a main, narrow peak as well as a broad, fainter component that is analogous to the spectrum analyzed in detail by \citet{Sabbi03}.

An important difference is that, contrary to PSR J1740-5340A, the offset of the main peak by $>$100 km s$^{\textrm{-1}}$ from the radial velocity of the companion for J1417 implies this emission feature does not likely originate in the stellar atmosphere, echoing our previous discussion. Furthermore, in the spectra we show nearest to opposition and conjunction of the secondary (panels III \& IV, respectively), the profiles show markedly less pronounced double-peak morphology; while one of the H$\alpha$ peaks is very broad and enhanced, the other virtually disappears. This odd Doppler-shifted ``flaring'' has not been observed in tMSPs in either state, but similar profiles in the black hole binary V404 Cyg have been explained as being due to an outflow from the central regions of the binary that absorbs one of the components \citep{Hynes02}. 

\subsubsection{The Source of H$\alpha$}

Given all the evidence, we suggest that the strange phenomenology in H$\alpha$ is likely not due to emission from an accretion disk, but instead comes from some combination of material flowing to the inner Lagrange point, hot gas swept out beyond the companion's orbit by the pulsar's radiation pressure, and an intrabinary shock. H$\alpha$ photons coming from an intrabinary shock or giant winds would be slightly offset from the orbital motion of the secondary, and display a very broad range of velocities, as we observe in nearly all our spectra. In both these circumstances, the emitting gas would slightly trail the secondary in its orbit. Matter streaming to the inner Lagrange point, which is then pushed away by the pulsar wind, would also generally track the orbital motion of the secondary, but always be slightly shifted, trailing the companion and moving roughly perpendicular to the line connecting the primary and secondary \citep[see Figure 2 in][]{Burderi02}. In general, at binary phases $\phi \sim 0.25$--0.75, these H$\alpha$ emitting photons would be blueshifted with respect to the secondary, and vice versa at phases $\phi$=0.0--0.25 \& 0.75--1.0. As described at the beginning of this section, this is precisely the global trend we see in the profiles across all epochs (Figure~\ref{fig:drawing}).

Whether this qualitative scenario can reproduce the detailed profiles at all epochs, as well as the X-ray light curve of the binary, remains to be tested with improved data and future detailed modeling.

\section{Conclusions}
\label{sec:discussion}
We have presented multiwavelength follow-up observations of the MSP binary J1417, a system likely in the late stages of the standard MSP recycling process that began after the secondary evolved off the main sequence and filled its Roche Lobe. 
Recently, \citet{Swihart17} reported on the discovery of a compact binary assocated with the unidentified \emph{Fermi} source 2FGL J0846.0+2820. This system, similar to J1417, also has an inferred heavy neutron star primary with a giant secondary in a wide 8.1 d orbit. Although no radio pulsations have yet been discovered in 2FGL J0846.0+2820, given the clear differences between these systems and the ``redback" and ``black widow" subclasses of MSP binaries, we have termed these systems with giant companions \emph{huntsman} binaries after the large huntsman spider.

From the perspective of the pulsar, the solid angle subtended by the secondary in J1417 is 1.3--1.4\%, comparable to the value for typical redback systems with unevolved donors \citep{Roberts15}. However, if some of the X-ray emission comes from a shock close to the secondary, then the effective solid angle is larger. While the light curve itself shows no evidence for irradiation (\S~\ref{sec:optLCs}), consistent with the high luminosity of the evolved secondary, the evidence for a wind from the optical spectroscopy shows that a substantial amount of material is leaving the system. This material is likely responsible for the difficulty in detecting pulsations from the MSP, even compared to typical redbacks \citep{Camilo16}. It is possible that similar huntsman systems remain undetected for this same reason.

The radio flux densities for J1417 are consistent with those expected for a MSP with modest scintillation, excepting the few VLA epochs where the pulsar is not detected. As it happens, these epochs were at superior or inferior conjunction, which is when the pulsar was also not detected in timing observations \citep{Camilo16}. This would be consistent with a scenario in which the radio emission was absorbed rather than scattered by the material in the system. On the other hand, both ATCA detections were obtained at or just after inferior conjunction.
Hence there may be both stochastic and orbital components to the variations in eclipsing material, consistent with the conclusions we draw for the H$\alpha$ variations in \S\ref{sec:Halpha}.

\citet{Strader15} pointed out that the 3FGL \emph{Fermi}-LAT spectrum of J1417 was a power-law with no evidence for curvature. This is unusual among MSPs, and is more similar to the $\gamma$-ray bright low-mass X-ray binaries 3FGL J1544.6--1125 and 3FGL J0427.9--6704 \citep{Acero15, Bogdanov16, Strader16}. If this power-law spectrum is confirmed in the forthcoming 4FGL \emph{Fermi}-LAT catalog, then it could suggest another source of high-energy emission in the system beyond simply the pulsar magnetosphere.

The new X-ray observations have allowed a better comparison of the X-ray properties of J1417 to known redbacks and tMSPs. At our new optically inferred distance (3.1 $\pm$ 0.6 kpc), $L_X = 0.7-1.4 \times 10^{33}$ erg s$^{-1}$, which is more typical of tMSPs in their accretion-powered state and brighter than all known redbacks in their pulsar state. If there is no accretion disk, as we suspect, it is likely that the bulk of this X-ray emission is coming from a strong intrabinary shock near the surface of the secondary. The tidally locked giant could generate a substantial magnetic field that may facilitate the formation of a luminous shock.

The most important new data to interpret J1417 would be an X-ray light curve complete in orbital phase, elucidating the nature of an intrabinary shock and allowing modeling of the H$\alpha$ emission in this context. Forthcoming \emph{Gaia} data releases should address systematics in the parallax measurements and definitively confirm the unusually high X-ray luminosity of the system. The unusual properties of this huntsman binary motivate ongoing searches for similar pulsar binaries.

\section*{Acknowledgements}

We extend a special thanks to Jerry Orosz for providing us with the latest version of the ELC code. We gratefully acknowledge support from NASA grants Chandra-GO6-17035X, 80NSSC17K0507, and NNX15AU83G, and NSF grant AST-1714825. J.S.~acknowledges support from the Packard Foundation. CH acknowledges support from NSERC. JCAM-J is the recipient of an Australian Research Council Future Fellowship (FT140101082). The National Radio Astronomy Observatory is a facility of the National Science Foundation operated under cooperative agreement by Associated Universities, Inc. The Australia Telescope Compact Array is part of the Australia Telescope National Facility which is funded by the Australian Government for operation as a National Facility managed by CSIRO.

\bibliography{report}

\end{document}